\documentclass[useAMS,usenatbib]{mn2e}
\usepackage{amsmath,amstext,amsgen,amsbsy,amsopn,amsfonts,theorem}

\usepackage[pdftex]{graphicx}
\usepackage{xspace}
\usepackage{amsmath}
\usepackage{hyperref}
\usepackage{framed}
\usepackage{txfonts}
\usepackage{epstopdf}
\usepackage{gensymb}

\usepackage{subcaption}

\usepackage[normalem]{ulem}

\makeatletter
\def\mr@ignsp#1 {\ifx\:#1\@empty\else #1\expandafter\mr@ignsp\fi}%
\newcommand{\multiref}[1]{\begingroup
\xdef\mr@no@sparg{\expandafter\mr@ignsp#1 \: }%
\def\mr@comma{}%
\@for\mr@refs:=\mr@no@sparg\do{\mr@comma\def\mr@comma{,}\ref{\mr@refs}}%
\endgroup}
\makeatother

\newcommand{\hypref}[2]{\ifx\href\asklfhas #2\else\href{#1}{#2}\fi}
\newcommand{\Secref}[1]{Section~\multiref{#1}}

\newcommand{\Figref}[1]{Figure~\multiref{#1}}
\newcommand{\figref}[1]{Fig.~\multiref{#1}}
\renewcommand{\eqref}[1]{(\multiref{#1})}

\newcommand{\eq}[1]{\begin{align}#1\end{align}}

\title[Biased Milky Way mass]{Equilibrium models of the Milky Way mass are biased high by the LMC}

\author[D. Erkal, V. A. Belokurov, \& D. L. Parkin]
  { Denis Erkal$^{1}$\thanks{d.erkal@surrey.ac.uk}, Vasily A. Belokurov$^{2}$, and Daniel L. Parkin$^{1}$ \\
  $^1$Department of Physics, University of Surrey, Guildford GU2 7XH, UK \\
  $^2$Institute of Astronomy, University of Cambridge, Madingley Road, CB3 0HA, Cambridge, UK
   }

\begin{document}

\label{firstpage}

\maketitle

\begin{abstract}
Recent measurements suggest that the Large Magellanic Cloud (LMC) may
weigh as much as 25\% of the Milky Way. In this work we explore how
such a large satellite affects mass estimates of the Milky Way based
on equilibrium modelling of the stellar halo or other tracers. In
particular, we show that if the LMC is ignored, the Milky Way mass is
overestimated by as much as 50\%. This bias is due to the bulk motion
in the outskirts of the Galaxy's halo and can be, at least in part,
accounted for with a simple modification to the equilibrium modelling. Finally, we show that the LMC has a substantial effect on the orbit Leo I which acts to increase its present day speed relative to the Milky Way. We estimate that accounting for a $1.5\times10^{11} M_\odot$ LMC would lower the inferred Milky Way mass to $\sim10^{12} M_\odot$.
\end{abstract}

\begin{keywords}
 Galaxy: kinematics and dynamics, Galaxy: evolution, galaxies: Magellanic Clouds
\end{keywords}

\section{Introduction}

The Large Magellanic Cloud (LMC) is the brightest satellite of the Milky Way and has been known since antiquity \citep[e.g.][]{alsufi}. The first suggestion that the LMC could have a significant effect on our Galaxy was proposed by \cite{kerr_1957} and \cite{burke_1957} based on the observations of the deformed atomic hydrogen disk in the Milky Way. However, calculations in those works based on the mass of the LMC at the time showed that the LMC was unlikely to explain the deformation. \cite{avner_king_1967} followed this up with a more general exploration of the effect of the LMC. Their discussion mostly focused on how the LMC could torque and twist the disk of the Milky Way assuming it was on a relatively circular orbit, although they also included a prescient discussion of whether or not the LMC was bound to the Small Magellanic Cloud (SMC) or to our Galaxy.

More recent work has shown that the Magellanic Clouds are likely bound to each other and are on their first approach to the Milky Way \citep{kallivayalil_2006_LMCSMC_orbit, besla_etal_2007,kallivayalil_etal_2013}. Alongside this, a number of works have shown that the LMC has a large total mass, $M_{\rm LMC} \sim 10^{11}-2.5\times10^{11} M_\odot$, based on abundance matching \citep[e.g.][]{moster_etal_2013}, assuming the SMC was originally bound to the LMC \citep{kallivayalil_etal_2013}, using the timing argument with Andromeda combined with the nearby Hubble flow \citep{penarrubia_lmc}, perturbations to the Milky Way disk \citep[][]{laporte+2018,gardner+2020}, quantifying the effect of the LMC on Orphan stream \citep{orphan_modelling}, and modelling the satellites of the LMC \citep{lmc_sats}. 

These high LMC masses can induce a substantial reflex motion in the Milky Way \citep{gomez_et_al_2015} and fits to the Orphan stream predicted this could be as large as $\sim 50$ km/s \citep{orphan_modelling}. This reflex motion should be most apparent beyond $\sim 30$ kpc where the orbital periods of stars are longer than the infall time of the LMC. Along these lines, \cite{lmc_wake} studied the effect of the LMC on the stellar halo of the Milky Way and predicted that there should be an over-dense wake behind the LMC as well as substantial non-equilibrium motion in the outskirts of our Galaxy.  \cite{pisces_LMC_wake} showed that the Pisces over-density \citep{Watkins2009} was consistent with this wake both in 3d shape and radial velocity. \cite{petersen_penarrubia} also simulated the infall of the LMC and found similar results. Thus, multiple lines of evidence suggest that the LMC has had a large effect on our Galaxy, making it significantly out of equilibrium.

In this work we will examine how the disequilibrium of our Galaxy
affects our ability to measure its mass. In particular, we will use
the mass estimator from \cite{watkins_etal_2010} which assumes that
our Galaxy is in equilibrium. This estimator has been used in a number
of recent works to measure the mass of our Galaxy out to radii where
these non-equilibrium effects are significant
\citep[e.g.][]{sohn+2018,watkins_etal_2019,fritz_2020}. We will also explore how
this reflex motion affects the mass estimate based on Leo I
\citep{boylan-kolchin_etal_2013}. This paper is organized as
follows. In \Secref{sec:LMC_effect} we will explore the effect of the
LMC on our Galaxy and in particular how it biases the mass
estimator. In \Secref{sec:discussion} we search for the predicted
reflex motion using satellites of the Milky Way, explore how this
reflex motion affects the motion of Leo I, and conclude.

\section{Effect of LMC on the Milky Way's halo} \label{sec:LMC_effect}

In order to simulate the effect of the LMC on a tracer population
around the Milky Way, we use a suite of simulations with a variety of LMC masses. The fiducial set of simulations are identical to those described in
\cite{pisces_LMC_wake}. The simulations in that work are fast since
the Milky Way and LMC are not resolved with full $N$-body
dynamics. Instead, they are modelled as single particles sourcing
their respective potentials. The Milky Way is modelled using the \texttt{MWPotential2014} from \cite{bovy_galpy} where the bulge is replaced by an equal mass Hernquist \cite{hernquist_1990} profile for speed. The LMC is modeled as a Hernquist profile. In this framework, the LMC is rewound
from its present day position, the stellar halo is initialized as a
population of tracer particles, and the system is evolved forward to
the present. We assume that the Sun is located at a distance of 8.122 kpc from the Galactic center \citep{2018A&A...615L..15G}.

\begin{figure*}
\centering
\includegraphics[width=0.99\textwidth]{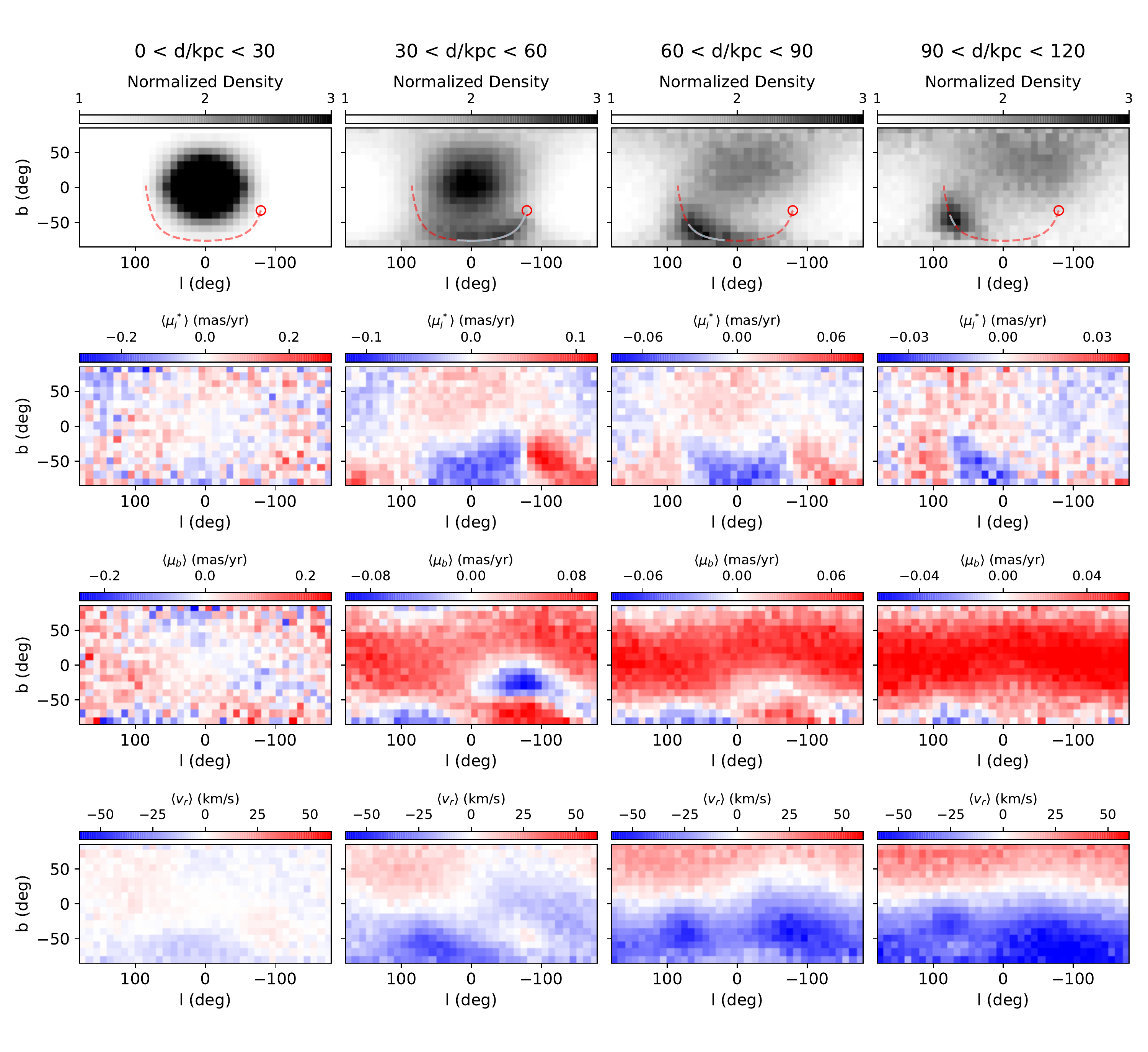}
\caption{Effect of a $1.5\times10^{11} M_\odot$ LMC on the Milky Way stellar halo in various distance bins. Each column shows a different range of distances (shown at the top of the plot). \textbf{Top row} shows the normalized stellar halo density. The red circle shows the present-day position of the LMC, the dashed-red curve shows the past orbit of the LMC, and the light blue curve shows the section of the LMC's orbit in the given distance bin. There is a clear overdensity in the outer stellar halo which traces the past orbit of the LMC. \textbf{Second row} shows the mean reflex-corrected proper motion $\mu_l^*$. In the inner most distance slice, there is very little structure. However, in the more distant slices there are clear patterns in the proper motion related to the past orbit of the LMC. \textbf{Third row} shows the mean reflex-corrected proper motion $\mu_b$. Beyond 30 kpc, the mean proper motion is positive over most of the sky, showing that the stellar halo is moving upwards relative to the inner part of the Galaxy. \textbf{Bottom row} shows the mean reflex-corrected radial velocity. Beyond 30 kpc, there is a clear dipole with the halo in the Southern hemisphere having a negative radial velocity (i.e. approaching the Sun) and the halo in the Northern hemisphere having a positive radial velocity (i.e. receding from the Sun). This also shows that inner part of the Galaxy is moving downwards with respect to the outer part of the stellar halo. These results are qualitatively similar to the results in \protect\cite{lmc_wake}.}  \label{fig:summary}
\end{figure*}

\Figref{fig:summary} shows the effect of a $1.5\times10^{11} M_\odot$
LMC on the stellar halo of the Milky Way\footnote{The simulation output is available at \url{https://doi.org/10.5281/zenodo.3630283}}. This demonstrates that this
technique gives broadly similar answers to \cite{lmc_wake} which
studied the motions of both the LMC and Milky Way with $N$-body
simulations. Comparing the two approaches it is clear that both the
``local wake" (aligned with the LMC's past orbit) and the ``global
wake" (mostly in the Northern hemisphere) are reproduced. This is due
to the fact that the simulations in \cite{pisces_LMC_wake} accounted
for the direct effect of the LMC on the stellar halo, as well as the
reflex motion of the Milky Way in response to the LMC. However, two
key aspects missing from our simulations are the deformation of the
Milky Way and LMC in response to each other and any resonances in the
Milky Way's halo. Given that \Figref{fig:summary} closely resembles
the results of \cite{lmc_wake}, these effects do not seem to be
important for the bulk properties of the stellar halo.

One key result highlighted in \Figref{fig:summary} is that the LMC
induces a motion of the inner region of the Milky Way, within roughly
30 kpc, with respect to the outer part of the Galaxy. This streaming
motion is mostly in the downwards (-$z$) direction. This is apparent
in the third and fourth rows of \Figref{fig:summary} which show that
the distant stellar halo is moving upwards relative to the inner parts
of the Galaxy. This effect was predicted in \cite{orphan_modelling}
which measured the mass of the LMC based on its effect on the Orphan
stream. That work argued that the orbital timescales are short in the
inner part of the Galaxy and thus these stars can respond
adiabatically to the LMC, while stars in the outskirts of the Galaxy,
where the orbital timescales are longer, do not respond coherently,
thus giving rise to a bulk motion. Since the LMC's past orbit has most
recently been below the Milky Way, the Cloud can be seen as pulling
the inner part of the Milky Way downwards. This effect was also seen
in the simulations in \cite{lmc_wake}.

This streaming motion suggests that
equilibrium modelling will likely be biased in the presence of the
LMC. To quantify this systematic error, we focus on the mass
estimator from \cite{watkins_etal_2010} which makes use of the tracers
3d velocity. This has been used in several recent works
\citep[e.g.][]{sohn_etal_2018,watkins_etal_2019,fritz_2020} but due to
the streaming motion relative to the outskirts of our Galaxy, we
expect that any method which assumes dynamical equilibrium will be
biased.  This estimate is given by
\eq{ M = \frac{\alpha+\gamma-2\beta}{3-2\beta} \frac{r_{\rm out}^{1-\alpha}}{G} \langle v^2 r^\alpha \rangle \label{eq:watkins}, }
where $\alpha$ is the power-law slope of the potential
(i.e. $\psi(r)\propto r^{-\alpha}$), $\beta$ is the anisotropy of the
tracer population, $\gamma$ is the power-law slope of the tracer
density (i.e. $\rho \propto r^{-\gamma}$), $r$ is the galactocentric
distance to each tracer, $v^2$ is the 3d speed of each tracer relative
to the Galaxy, and $r_{\rm out}$ is the radius of the outermost
tracer.

\begin{figure}
\centering
\includegraphics[width=0.49\textwidth]{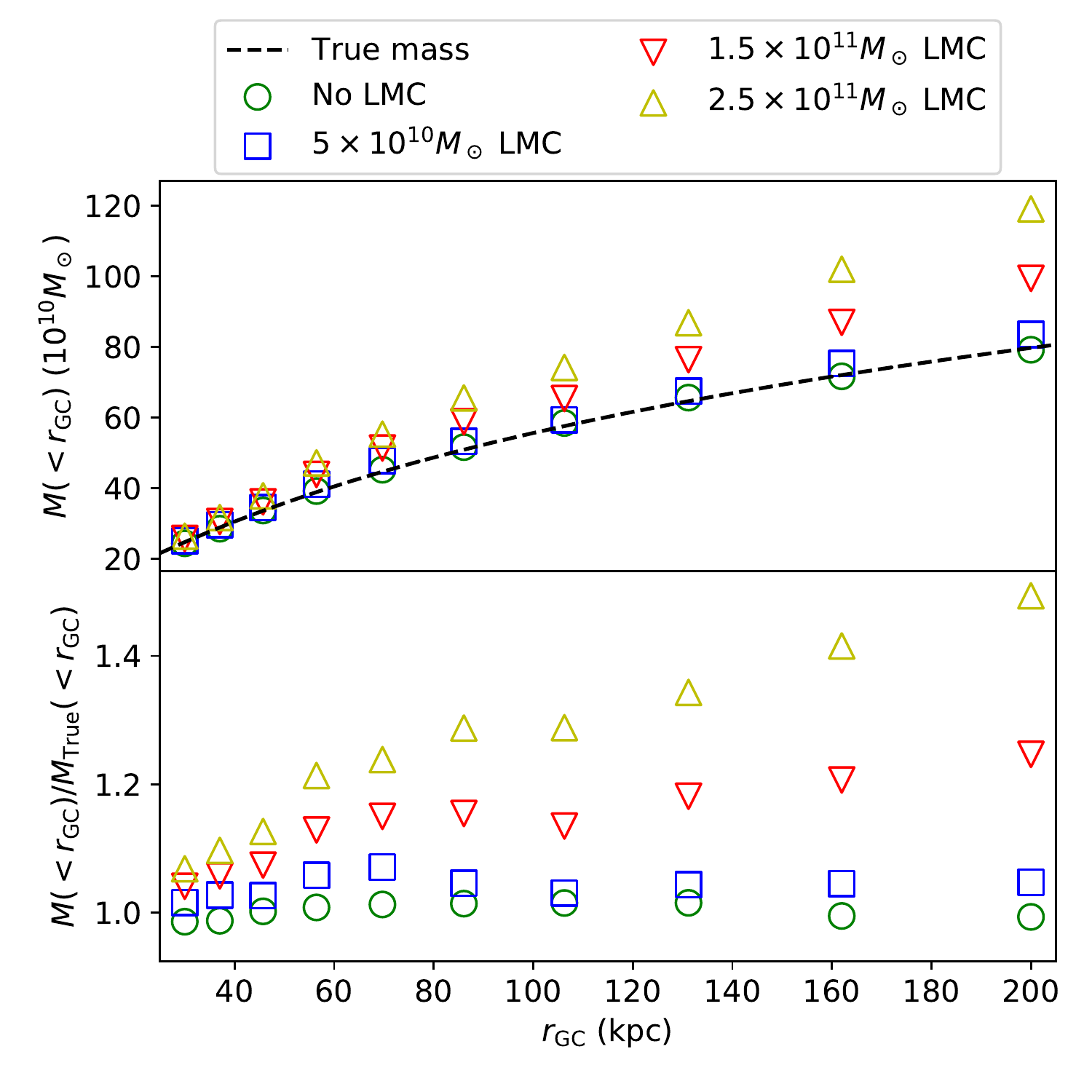}
\caption{Mass estimator applied to simulated Milky Way in the presence of the LMC. \textbf{Top panel} shows the mass estimator applied to four simulations with various LMC masses. The simulation with no LMC matches the true mass quite well while as the LMC mass is increased, the inferred mass is increasingly biased. As expected, the estimates converge to the true value within $\sim$ 30 kpc since the Milky Way does not have any bulk motion within this region. \textbf{Bottom panel} shows the inferred mass divided by the true mass, i.e. the size of the bias. For the most massive LMC considered here, $2.5\times 10^{11} M_\odot$, the inferred mass is $\sim$ 50\% higher than the true mass at 200 kpc. However, even at more modest radii of $\sim$ 50 kpc, this bias can reach 20\%. }  \label{fig:m_est}
\end{figure}

We apply this mass estimator to the simulated stellar haloes from
\cite{pisces_LMC_wake}. In particular, we use the $5\times10^{10}
M_\odot$, $1.5\times10^{11} M_\odot$, $2.5\times10^{11} M_\odot$ LMC
runs and the run with no LMC. First, as a reference, we use
Equation~\ref{eq:watkins} to measure the MW mass in the simulation
without the LMC. For this stellar halo, we expect that the estimator
in \eqref{eq:watkins} will recover the true mass profile. We break the
stellar halo into radial bins from 30 kpc to 200 kpc. For each bin, we
use a bin width which is $\sim$19\% of the radius so that the mass
estimate is as accurate as possible. This results in 10 bins in the
range 30-200 kpc. If we use significantly larger bins, then the
power-law slope of the potential and tracer density can change
significantly within each bin, making the estimator less precise. The
results are shown in \Figref{fig:m_est}. The green circles show the
mass estimator applied to the simulation with no LMC. As expected,
this faithfully reproduces the true mass distribution of the simulated
Milky Way.

Next we consider the simulations with three different LMC masses. The
resulting mass profiles are also shown in \Figref{fig:m_est} using
different color symbols. As the LMC mass is increased, the Milky Way
mass is progressively overestimated. Indeed, at 200 kpc, this can
result in an up to $\sim$ 50\% overestimate of the Milky Way
mass. Interestingly, all of the measurements converge to the true
Milky Way mass within $\sim 30$ kpc since within this region there is
no significant bulk motion and the Milky Way is effectively in
equilibrium.

\begin{figure}
\centering
\includegraphics[width=0.49\textwidth]{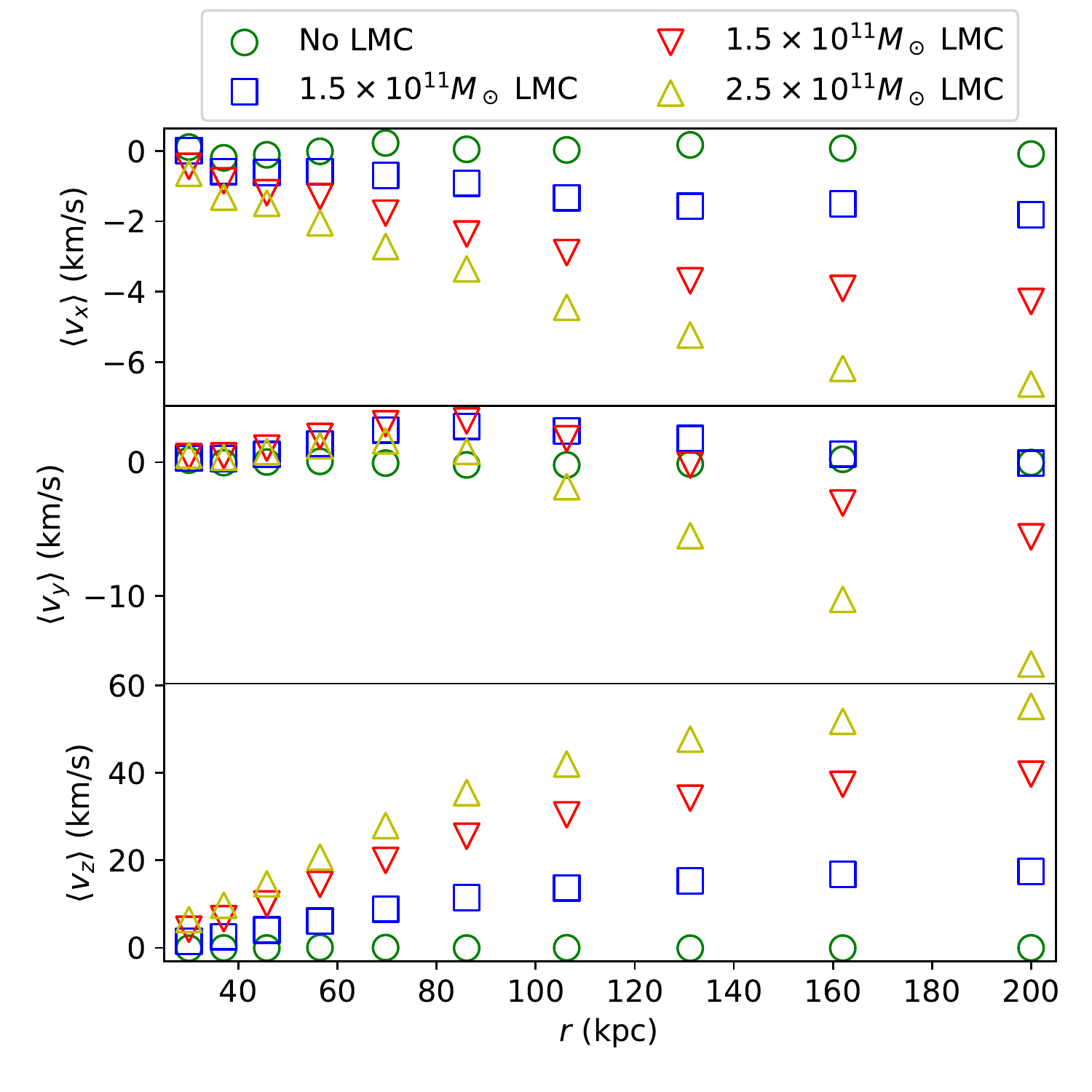}
\caption{Magnitude of mean velocity in radial shells in the presence of various mass LMCs. If the LMC is not included, the mean velocity in all shells is close to zero, as expected if the system is in equilibrium. However, as the LMC mass is increased, the mean velocity beyond $\sim 30$ kpc increases to nearly $60$ km/s in the case of the $2.5\times10^{11} M_\odot$ LMC. The velocity is mainly in the $+z$ direction (i.e. upwards): over the radial range considered here, the velocity is always within $\sim 18^\circ$ of the $+z$ direction. This large bulk velocity shows that the outer parts of the stellar halo are significantly out of equilibrium due to the LMC. Note that the y-range of each panel is different.}  \label{fig:mean_v}
\end{figure}

In \Figref{fig:mean_v} we explore the bulk velocity induced by the
LMC's in-fall by computing the tracer mean velocity in radial
shells. If the LMC is not included, the mean velocity is close to zero
as expected since the stellar halo is in equilibrium. However, as the
LMC mass is increased, the mean velocity grows significantly in the
outer parts of the halo ($r > \sim 30$ kpc). The bulk motion of the
distant stellar halo is mostly in the upwards (i.e. $+z$
direction). This is due to the fact that in the recent past, the LMC's
orbit has taken it below the plane of the Milky Way. As the Cloud
passes its peri-centre, the short orbital timescales in the inner part
of the Milky Way allow it to respond coherently, while the timescales
in the outer part are too long. Thus, the inner part of the Milky Way
is accelerated downwards relative to the outer parts as argued in
\cite{orphan_modelling}.

\begin{figure}
\centering
\includegraphics[width=0.49\textwidth]{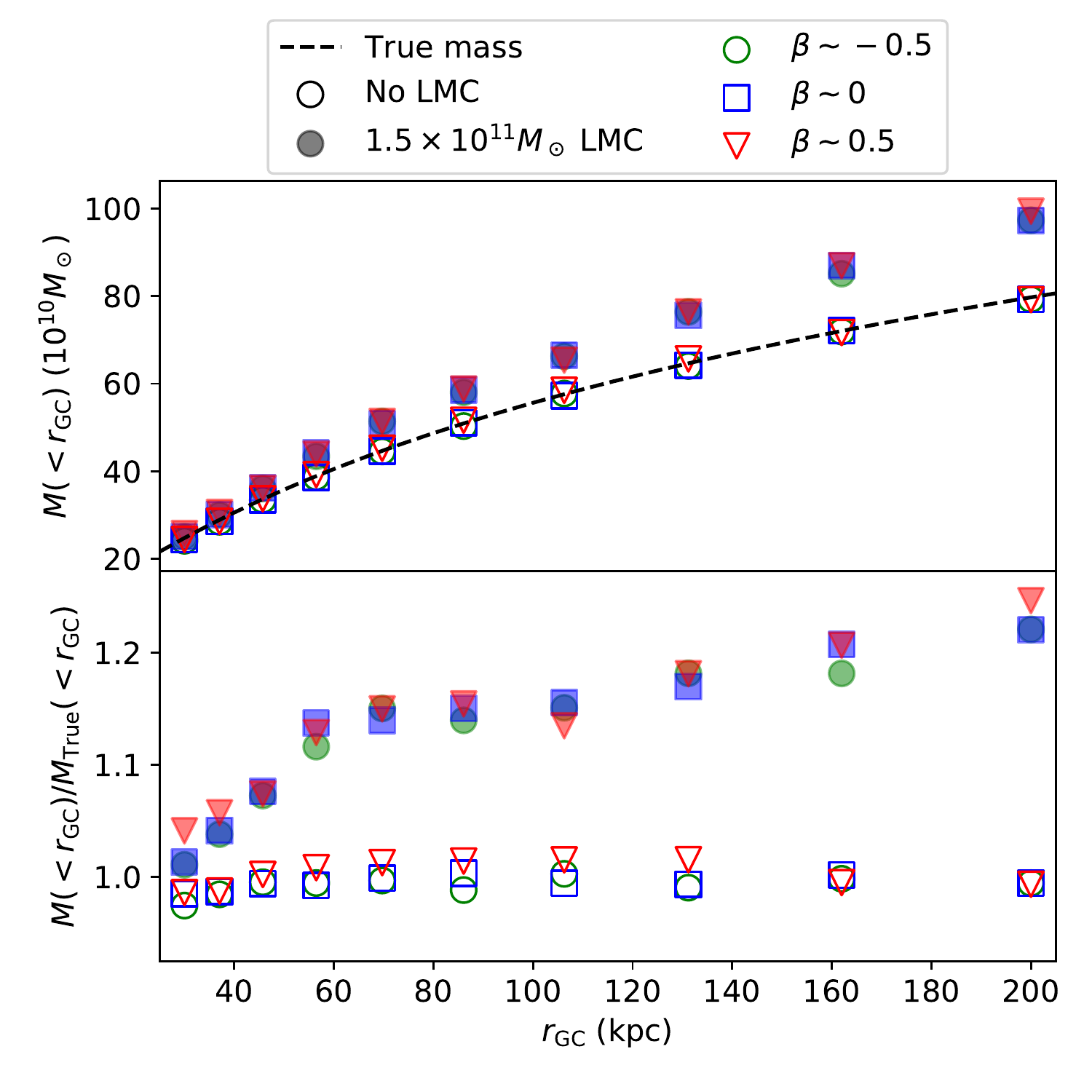}
\caption{Mass estimator applied to stellar haloes with various anisotropies in the presence of the LMC. \textbf{Top panel} shows the mass estimator compared to the true mass and the \textbf{bottom panel} shows the mass estimator divided by the true mass. Since the anisotropy does not significantly affect the mass estimate, the bias induced by the LMC should be present in all tracer populations.}  \label{fig:m_est_ani}
\end{figure}

The simulations in \cite{pisces_LMC_wake} were tailored to
be similar to the stellar halo since they have an anisotropy of $\sim
0.5$ in agreement with recent measurements in the Galaxy \citep[see
  e.g.][]{Lancaster2019}. In order to study how other tracer
populations, i.e. globular clusters or dwarf galaxies, are affected we
run two additional simulations. For these we have an anisotropy of
$\sim 0$ and $\sim$ -$0.5$ to investigate how changing the anisotropy
affects our results. As in \cite{pisces_LMC_wake}, the initial conditions are generated using \textsc{agama} \citep{agama} with the \texttt{DoublePowerLaw} distribution functions from \cite{posti_etal_2015}. For an anisotropy of $\sim$ 0, we use \texttt{norm=1.5e10},
\texttt{j0=500}, \texttt{slopeIn=0}, \texttt{slopeOut=3.5},
\texttt{coefJrOut=1.175}, \texttt{coefJzOut=0.9125},
\texttt{jcutoff=1e5}, \texttt{cutoffStrength=2} and for an anisotropy of $\sim$ -$0.5$, we use \texttt{norm=1.5e10},
\texttt{j0=500}, \texttt{slopeIn=0}, \texttt{slopeOut=3.5},
\texttt{coefJrOut=1.48}, \texttt{coefJzOut=0.76},
\texttt{jcutoff=1e5}, \texttt{cutoffStrength=2}. For these different anisotropies, we only
consider an LMC mass of $1.5\times10^{11} M_\odot$. In
\Figref{fig:m_est_ani} we compare the mass estimator with three
different anisotropy values and see that the estimator is similarly
biased, independent of the anisotropy chosen. Thus, we should expect
that estimates of the Milky Way with any tracer will be strongly
affected.

Since a lot of the velocity structure in \Figref{fig:summary} appears
to be due to the inner part of the Galaxy moving downwards with
respect to the outer parts, we
propose a slightly modified version of the estimator from
\cite{watkins_etal_2010} which uses the velocity dispersion relative
to the mean velocity:
\eq{ M_{\bar v} = \frac{\alpha+\gamma-2\beta}{3-2\beta} \frac{r_{\rm out}^{1-\alpha}}{G} \langle (\mathbf{v}-\bar{\mathbf{v}})^2 r^\alpha \rangle \label{eq:watkins_mod}. }

\begin{figure}
\centering
\includegraphics[width=0.49\textwidth]{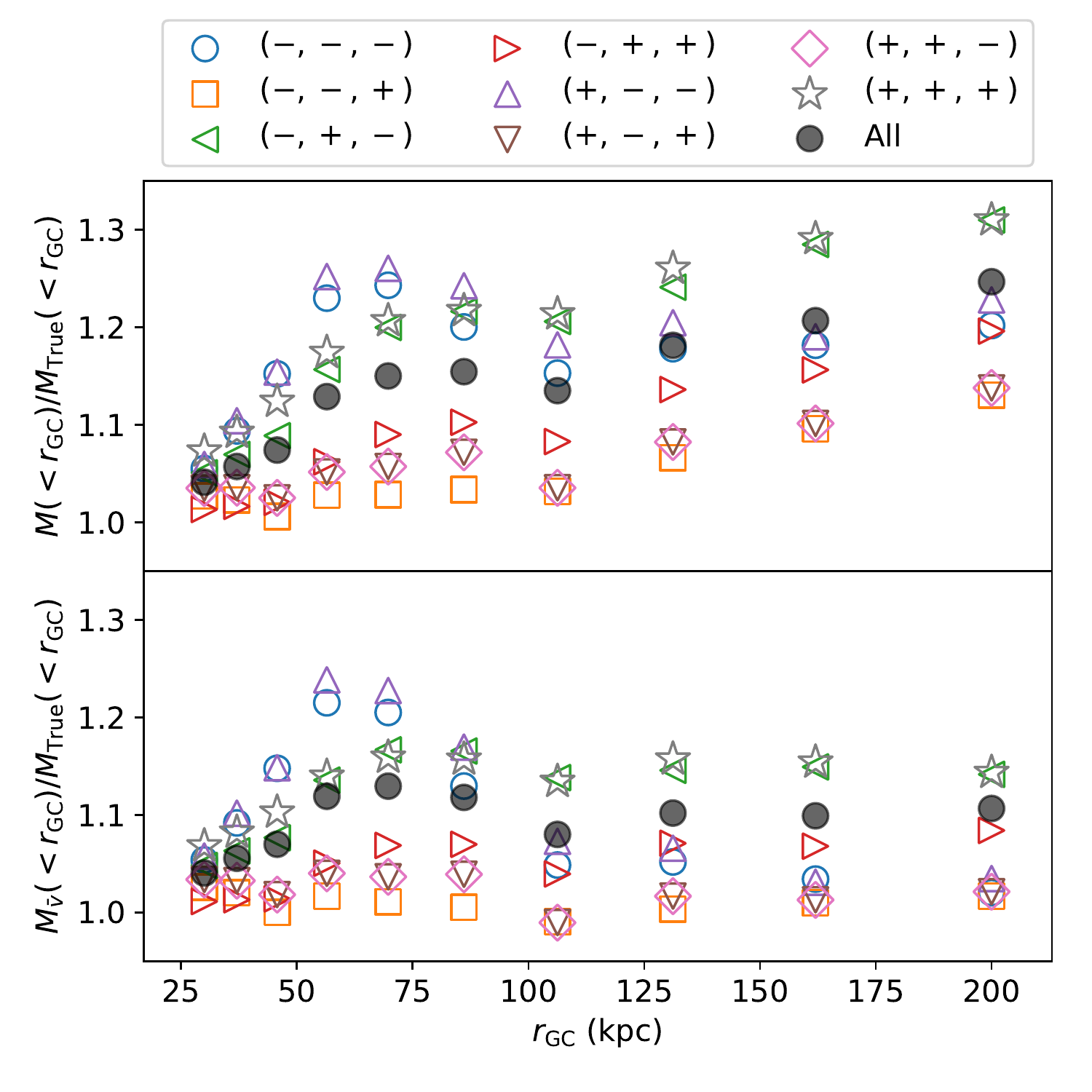}
\caption{Mass estimator applied on different octants in the presence of a $1.5\times10^{11} M_\odot$ LMC. \textbf{Top panel} shows the mass estimator applied on the octants of the Milky Way's standard galactocentric Cartesian coordinates. The octants are specified in the legend by the sign of the x,y,z coordinates respectively, e.g. $(-,-,-)$ means $x<0,y<0,z<0$. Given the velocity structure apparent in \protect\Figref{fig:summary} it is not surprising that the mass estimator applied to the various octants gives a different result.  While the majority of the most biased octants are in the Southern hemisphere (i.e. $z<0$), the octant $(+,+,+)$ is also significantly biased. \textbf{Bottom panel} shows the mass estimator which accounts for the bulk velocity of the tracers (see eq. \ref{eq:watkins_mod}). Several octants can now provide an unbiased estimate over a wide range of radii, especially octants $(-,-,+)$, $(+,+,-)$, $(+,-,+)$, and (slightly worse) $(-,+,+)$.    }\label{fig:m_est_oct}
\end{figure}

A comparison of this estimator versus the one in
Equation~\ref{eq:watkins} is shown in \Figref{fig:m_est_oct} for eight
different octants (marked according to their positive or negative
position along each of $X,Y, Z$ axes). Two of the best octants are
$(-,-,+)$ and $(+,-,+)$. These correspond to $l<0^\circ$ and
$b>0^\circ$. Referring back to \Figref{fig:summary}, this is
reassuringly the quadrant of the sky with the least structure in the
velocity.

\section{Discussion and conclusions} \label{sec:discussion}

\subsection{Search for velocity shift in the observations}

In \Figref{fig:summary} we showed that the LMC has a large effect on
the outer parts of the Milky Way. One of the main results is that the
outer regions of the Milky Way are moving upwards relative to the
inner regions. In order to test this, we use a sample of 33 globular
clusters and dwarf galaxies with galactocentric radii larger than 30
kpc. The data for the globular clusters come from \cite{vasiliev_GCs}
and references therein. Since no distance error is provided, we assume
an error of 5\% for each globular cluster which corresponds to a
distance modulus error of 0.1 mag
\citep[e.g.][]{gratton+2003,correnti+2018}. The data for the
ultra-faint dwarfs come from
\cite{simon_umaI_umaII_comB,koposov_booI,willman_wilI,kirby_seg2,martin_dra2,walker_tuc2_gru1,simon_tuc3,kirby_tri2,simon_gaia_pms,pace_li_pms,antlia2}. For
the classical dwarfs, we use the observed values from \cite{gaia_pms}
as well as proper motions for Leo I and Leo II from
\cite{sohn_etal_2013} and \cite{leoii_pm} respectively. In order to
avoid any obvious clustering, we exclude the dwarfs associated with
the LMC, including the LMC and SMC
\citep{kallivayalil_etal_2018,lmc_sats,patel_lmc_sats}.

\begin{figure}
\centering
\includegraphics[width=0.49\textwidth]{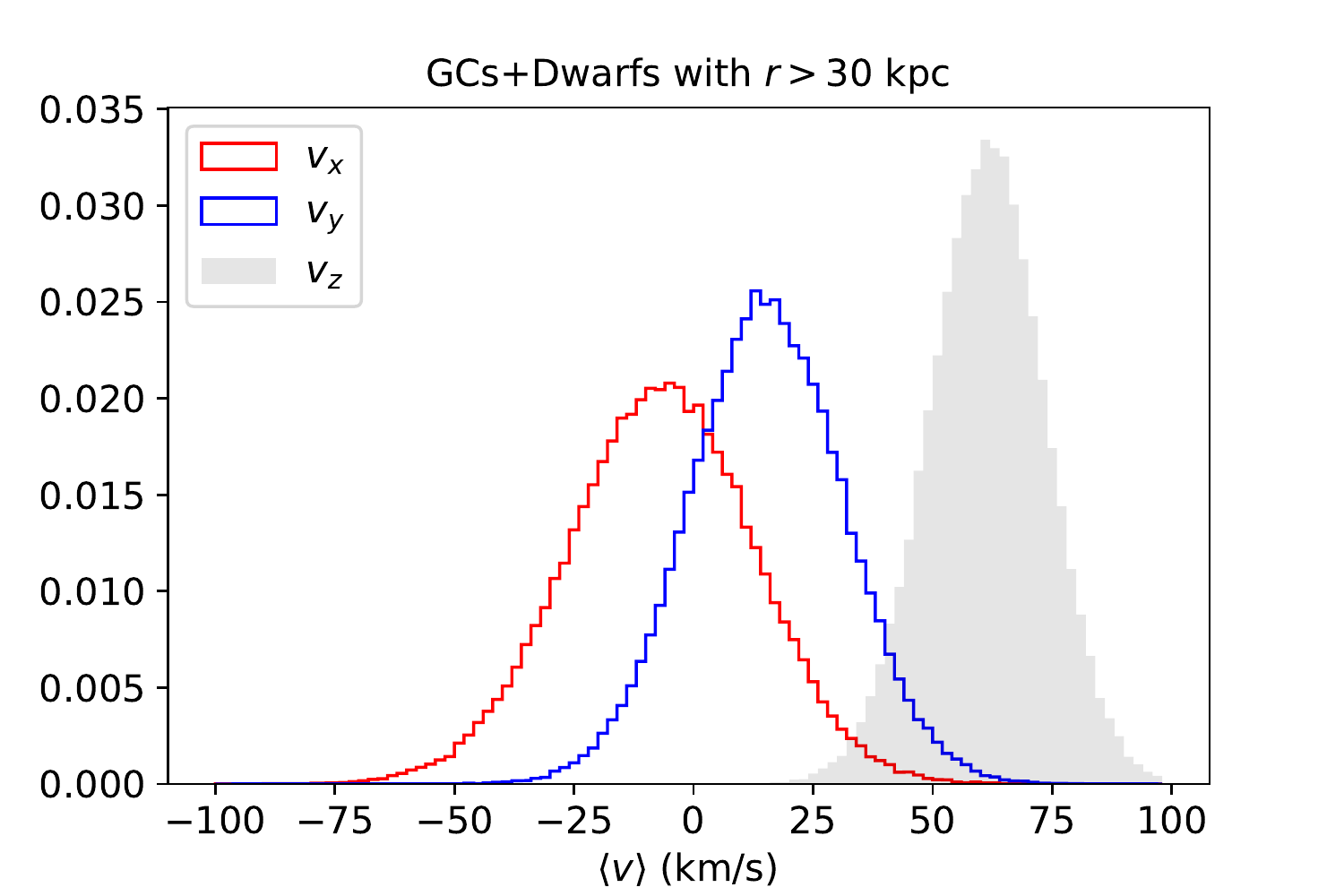}
\caption{Mean velocity in sample of 33 globular clusters and dwarfs with galactocentric distances larger than 30 kpc. This shows that these satellites have a substantial velocity shift relative to the Milky Way. This could be due to the reflex motion induced by the LMC or due to substrcture and phase-space correlation amongst the satellites. }\label{fig:mean_v_obs}
\end{figure}

For this sample, we then make 100,000 Monte Carlo realizations of
their cartesian velocities (given the observables and their
uncertainties) and compute the mean of each cartesian velocity. We note that the cut at 30 kpc is made after each Monte Carlo realization, i.e. some satellites only contribute in a fraction of the realizations. The distributions of these means are shown in \Figref{fig:mean_v_obs}
where we see that while $v_x$ and $v_y$ have means which are
consistent with zero, the mean of $v_z$ is significantly non-zero and
positive. This is in line with the predicted effect of the
LMC (see \figref{fig:mean_v}). However, we note that since satellites are known to arrive to the
MW in associations, it is possible that this signal is due to recently
accreted groups of dwarf galaxies which have not yet phase-mixed in
their orbits around the Milky Way. Future observations of the stellar halo will
verify whether this signal is due to the LMC. 

Along similar lines, we note that \cite{gilbert_etal_2018} has shown that there is a
significant velocity offset between the stellar halo and disk in
Andromeda (see Fig. 7 of that work). Such an offset could arise from
an interaction between Andromeda and a large satellite as in this
work. Thus, this may be due to the recent merger proposed by
\cite{hammer_etal_2018}.

\subsection{Leo I}

The large relative speed of the Leo I dwarf galaxy relative to the
Milky Way \citep{sohn_etal_2013} has been used to constrain the mass
of the Milky Way \citep{boylan-kolchin_etal_2013} assuming that it is
bound to our Galaxy. However, since the outer parts of our Galaxy are
out of equilibrium, this relative speed already includes the
additional reflex motion imparted by the LMC. In order to assess the
impact of the LMC on Leo I we take two approaches. First, we estimate
this reflex motion using the fiducial simulations from
\Secref{sec:LMC_effect} with an LMC mass of $1.5\times10^{11}
M_\odot$. We take the mean velocity of all particles within 2 degrees
on the sky and 30 kpc along the line of sight from the currently
measured location of Leo I. Accounting for the reflex motion, the
relative speed of Leo I drops from $\sim 197$ km/s to $\sim168$ km/s.

Second, we integrate the orbit of Leo I in the presence of the LMC
using the machinery from \cite{lmc_sats}. Namely, we rewind Leo I back
in time for 5 Gyr (or until the LMC has an apocenter, whichever is
sooner) including the effect of a $1.5\times10^{11} M_\odot$ LMC. The
Milky Way and LMC potentials are the same as in
\Secref{sec:LMC_effect}. We Monte Carlo the present day positions and
velocities of the LMC and Leo I 10,000 times and compare the energy of
Leo I relative to the Milky Way at the present and 5 Gyr ago
(i.e. before the infall of the LMC). We find that the energy of Leo I
was substantially lower (i.e. it was substantially more bound) before
the infall of the LMC. In order to facilitate the comparison with
\cite{boylan-kolchin_etal_2013}, we convert this energy difference
into a change of the velocity of Leo I. Thus, we find that if Leo I
was observed at its current location before the infall of the LMC, it
would have had a speed of $\sim 169$ km/s relative to the Milky
Way. This is much lower than its present day relative speed of $\sim
197$ km/s.

Interestingly, both approaches give nearly the same result showing
consistently that a significant portion of Leo I's speed is due to the
LMC.  In terms of the results of \cite{boylan-kolchin_etal_2013}, this
$\sim15\%$ decrease in the speed is a slightly larger effect than
changing the Milky Way mass from $10^{12}M_\odot$ to
$1.5\times10^{12}M_\odot$ which results in a $\sim13\%$ reduction in
$v/v_{\rm vir}$. This suggests that if the analysis of
\cite{boylan-kolchin_etal_2013} was repeated accounting for a
$1.5\times10^{11} M_\odot$ LMC, the inferred Milky Way mass would be
close to $\sim10^{12} M_\odot$.

\subsection{Conclusions}

In this work we have shown that the LMC should push the outskirts of
our Galaxy substantially out of equilibrium. In particular, the
leading order effect of the LMC is that the inner parts of the Milky
Way nearly decouple from the region beyond $\sim 30$ kpc. Thus,
observations of populations beyond $\sim30$ kpc should show signs of
this near-bulk motion. Importantly, we demonstrated how this
disequilibrium affects models of matter distribution in the outer
parts of our Galaxy using the mass estimator of \cite{watkins_etal_2010}. The systematic bias in the tracer velocity
dispersion induces the mass bias which is always positive and can be
as large as $\sim 50\%$ depending on the mass of the LMC. We showed
that this bias depends on where the tracers are located and that
certain parts of the sky offer a substantially improved estimate. This
bias can also be reduced if the mean reflex motion is accounted
for. In a similar vein, we showed that the LMC significantly increases
the present-day speed of Leo I relative to the Milky Way and that if
this is accounted for, the Milky Way mass estimate of
\cite{boylan-kolchin_etal_2013} will be significantly lower.

Accounting for the reflex motion induced by the LMC may also bring
into closer agreement the different mass estimates for the Milky Way
\citep[e.g.][]{bland-hawthorn_gerhard_2016,wang+2019} which are made with
tracers at different radii and using different techniques. Based on the results of this work, the
estimates made with data in the outskirts of our Galaxy are likely
biased high due to the nearly bulk motion beyond $\sim 30$ kpc. Future
observations of the stellar halo with \textit{Gaia} DR3, as well as
upcoming radial velocity surveys like WEAVE and 4MOST, will allow us
to measure this bulk motion and determine how significant this effect
is.

\section*{Acknowledgements}

We thank Wyn Evans for helpful comments. We thank Eugene Vasiliev for help with using \textsc{agama}. This research made use of \textsc{ipython} \citep{IPython}, python packages \textsc{numpy} \citep{numpy}, \textsc{matplotlib} \citep{matplotlib}, and \textsc{scipy} \citep{scipy}. This research also made use of Astropy,\footnote{http://www.astropy.org} a community-developed core Python package for Astronomy \citep{astropy:2013, astropy:2018}.

\bibliographystyle{mn2e_long}
\bibliography{citations_lmc}

\end{document}